\documentclass[preprint,3p,onecolumn]{elsarticle}

\usepackage{times,bm}
\usepackage{subeqnarray}
\usepackage{enumerate,graphicx,amsmath,amssymb,latexsym,psfrag,epsfig}
\usepackage{theorem,fancyhdr,fancybox}
\usepackage{calrsfs}
\usepackage[normalem]{ulem}
\usepackage{lscape}

\newtheorem{theorem}{Theorem}
\newtheorem{lemma}{Lemma}

\newtheorem{definition}{Definition}

{\theorembodyfont{\upshape}}
{\theorembodyfont{\upshape}\newtheorem{remark}{Remark}}
{\theorembodyfont{\upshape}}

\def\diag{\mathop{\mathrm{diag}}\nolimits} 

\journal{Systems \& Control Letters}

\begin{document}

\begin{frontmatter}

\title{A Round-Robin Type Protocol for Distributed Estimation with $H_\infty$
  Consensus\tnoteref{t1}\tnoteref{t2}}

\author[label1]{V.~Ugrinovskii\corref{cor1}}
\ead{v.ougrinovski@adfa.edu.au}
\cortext[cor1]{Corresponding author}

\author[label2]{E.~Fridman}
\ead{emilia@eng.tau.ac.il}

\address[label1]{School of Engineering
   and Information Technology,  UNSW Canberra at the
 Australian Defence Force Academy, Canberra, Australia.}
\address[label2]{School of Electrical Engineering, Tel-Aviv
  University, Israel.}

\tnotetext[t1]{This research was supported under the Australian Research
  Council's Discovery Projects funding scheme (Project DP120102152) and by
  Israel Science Foundation (grant No 754/10). Part of this work was
  carried out during the first author's visit to the Australian National
  University.}

\tnotetext[t2]{A preliminary version of the results of this paper was
  presented at the 52nd IEEE CDC, Florence, Italy \cite{UFri1a}.}

\begin{abstract}
 The paper considers a distributed robust estimation problem over a network
 with directed topology involving continuous time observers. While
 measurements are available to the observers continuously, the nodes interact
 according to a Round-Robin rule, at discrete time instances. The results
 of the paper are sufficient conditions which guarantee a suboptimal $H_\infty$
 level of consensus between observers with sampled
 interconnections.
\end{abstract}

\begin{keyword}
Large-scale systems, distributed estimation, robust observers, consensus,
vector dissipativity, sampled-input systems, time-delay systems.
\end{keyword}

\end{frontmatter}

\section{Introduction}

The problem of distributed estimation is one of very active topics in the
modern control theory and signal processing literature. Interest in this
problem is motivated by a growing number of applications where a decision
about the observed process must be made simultaneously by spatially
distributed sensors, each taking partial measurements of the process.

When the process and measurements are subject to noise and disturbance,
robustness aspects of the problem come into prominence. In the past several
years, a number of results have been presented in the literature which
develop the $H_\infty$ control and estimation theory for distributed
systems subject to uncertain perturbations; e.g.,
see~\cite{DLG-2011,LDCH-2010,LJL-2008,NF-2009,U6,LaU1,U7}. In particular,
methodologies of distributed sampled-data $H_\infty$ filtering have been
considered, e.g., in~\cite{SWL-2011}. That reference emphasized several
distinctive 
aspects of realistic sensor networks, among them coupling between
sensor nodes through the information communicated between neighbouring
sensor nodes and the sampled nature of that coupling, which is dictated by
the digital communication technology. The latter feature of sampled data
networks is an important consideration in any network design, because the
amount of information that can be transmitted to/received at each node of
the network is constrained, due to data rate limitations of digital
communication channels.

In this paper we address some of the challenges specific to Round-Robin
type communication protocols.
The Round-Robin protocol is a commonly used protocol for information
transmission in networked control systems. It allows each node to communicate
with its neighbours intermittently, during scheduled time slots and is
known to lead to bandwidth savings. From a hybrid systems
perspective this protocol has been studied in details
in~\cite{NT-2004,HTWN-2010}. More recently, it has been considered in the
context of time-delay systems in~\cite{LFH-2012}, where an analysis of
exponential stability and $L_2$ properties of networked control systems
with Round-Robin scheduling was presented using a delay switching system
modeling. In this paper, we further develop this technique in the context
of robust distributed estimation with intermittent communications between
sensing nodes. The type of communication we consider is where the nodes
broadcast their information at every scheduled time instant to all nodes in
their vicinity, but they listen to only one node within their neighbourhood
at a time, according to the Round-Robin rule. For instance, this can be
achieved by encoding the transmitted information with a node-specific key. I.e.,
when node $i$ receives signals from multiple sources, it
extracts the information sent by node $j$ by utilizing the key of that
node, and continues doing this by rotating the keys.

The objective of this paper is to develop an
algorithm for synthesis of a Round-Robin type protocol for a network of
distributed observers, to allow this network to track dynamics of a linear
uncertain plant. Unlike
many existing approaches to distributed estimation, the salient feature of
our methodology is that individual estimators may not be able to track the
plant, if they rely
solely on their own measurements, because the plant may not be observable
from the node's measurements. This issue has recently been emphasized
in~\cite{U6,LaU1,U7} which demonstrated that consensus between sensors
plays a crucial role in enabling individual sensor nodes to overcome lack
of detectability and successfully track the plant. Necessary conditions on
the network to ensure the plant is detectable/observable by the network
have recently been presented in~\cite{U7b-journal,ZM-2008}.

The first contribution of this paper is a version of the protocol
of~\cite{LFH-2012} to be used with the distributed estimation schemes
proposed~\cite{U6,LaU1,U7}. We show that instead of continuously exchanging
information (the type of networks considered in those references),
the node observers can achieve the relative $H_\infty$ consensus objective
by exchanging information at certain sampling times, by polling one
neighbour at a time. It is assumed that sampling times are known and agreed
upon at each node. Technically, this would require all nodes to have
their clocks synchronized, e.g., by means of a network time protocol
\cite{Mills-1991}.

Our second contribution demonstrates that the Round-Robin design
of~\cite{LFH-2012} can be applied to derive a network of
non-switching observers. Of course, each observer periodically switches
between input channels, but the observer gains remain unchanged. This
is an important feature of our methodology to ensure its
scalability.  In a large network of distributed estimators, switching
of observer gains typically leads to a combinatorially complex scheduling
problem, and necessitates development of additional
tools to resolve this complexity; see~\cite{U7}. We show that these issues
are avoidable in observer networks of the type considered in this paper.

Our main result is a sufficient condition, expressed in the form of Linear
Matrix Inequalities (LMIs), from which filter and
interconnection gains for each node estimator can be computed, to ensure
the network of sampled data observers using switching communications
converges to the trajectory of the observed plant by achieving consensus
between the filters at every node. Conditions for consensus of multi-agent
systems using sampled data communications or systems communicating over
switching graphs are well known in the literature
\cite{OM-2004,CR-2010}. This includes an observation that the sampling
period has a significant effect on the system performance~\cite{HMH-2006}.
As the example presented in Section~\ref{example} illustrates, our
conditions allow to investigate the effects of intermittent sampled
communications on the system performance as well. At the same time, our
result provides a guarantee of the network consensus performance in the
presence of disturbances. Since consensus between observers is essential
for the network to be able to overcome observability/detectability
limitations of individual observers~\cite{U6,ZM-2008}, consensus
performance is seen as an important design consideration which the results
in this paper address.

As in~\cite{U6,LaU1,U7}, our methodology relies on
certain vector dissipativity properties of the
large-scale system comprised of the observers' error dynamics. However,
different form these references, to establish these vector dissipativity properties, we employ a novel class of generalized supply rates which reflect the
sampled-data nature of interconnections between observers. The general idea
behind introducing such generalized supply rates can be traced
to~\cite{LCD-2004} (also, see~\cite{LaU1}), but our proposal here makes use
of special properties of sampled signals. In the limit, when the maximum
sampling period approaches zero, these generalized supply rate vanish, and one
recovers the vector dissipativity properties of error dynamics established
in~\cite{U6}. Thus, the feasibility of the
conditions proposed for systems with continuously operating
interconnections can be used as a preliminary (but not conclusive) test
for the 
conditions proposed here. Our numerical example illustrates this point very
well, showing a negligible difference between the disturbance attenuation levels
obtained using the benchmark algorithm of~\cite{LaU1} and those obtained
using our conditions under small sampling rate. At the same time, the
feasibility of our conditions makes explicit the dependence of the proposed
algorithm on the sampling period. Being only sufficient conditions, our
conditions are potentially conservative, but the fact that the techniques used
in the derivation of these conditions showed substantial reduction of
conservatism in similar problems~\cite{LFH-2012} is encouraging.    

The paper is organized as follows. The problem formulation, along with the
graph theory preliminaries is presented in Section~\ref{Formulation}. The
main results of the paper are given in Section~\ref{main}. In
Section~\ref{example}, we discuss an illustrative example, and
Section~\ref{Concl} concludes the paper.

\paragraph*{Notation} Throughout the paper, $\mathbf{R}^n$ denotes a real
Euclidean $n$-dimensional vector space, with the norm $\|x\|\triangleq
(x'x)^{1/2}$; here the symbol $'$ denotes the transpose of a matrix or a
vector. $L_2[0,\infty)$ will denote the Lebesgue space of
$\mathbf{R}^n$-valued vector-functions $z(\cdot)$, defined on the time
interval $[0,\infty)$, with the norm $\|z\|_2\triangleq
\left(\int_0^\infty\|z(t)\|^2dt\right)^{1/2}$ and the inner product
$\int_0^\infty z_1(t)'z_2(t)dt$. $\otimes$ is the Kronecker product of
matrices, $\mathbf{1}_n\in \mathbf{R}^n$ is the column-vector of
ones. Also, $\det X$ is the determinant of $X$.

\section{The problem formulation}\label{Formulation}

\subsection{Graph theory}

Consider a filter network with
$N$ nodes and a directed graph topology $\mathcal{G} = (\mathcal{V},\mathcal{E})$;
$\mathcal{V}=\{1,2,\ldots,N\}$, $\mathcal{E}\subset \mathcal{V}\times
\mathcal{V}$ are the set of vertices and the set of edges, respectively.
The notation $(j,i)$ will denote the edge
of the graph originating at node $j$ and ending at node $i$. In accordance
with a common convention~\cite{OM-2004}, we consider graphs without
self-loops, i.e., $(i,i)\not\in\mathbf{E}$. However, each node is assumed
to have complete information about its filter and measurements.

For each $i\in \mathcal{V}$, we denote $\mathcal{V}_i=\{j:(j,i)\in
\mathcal{E}\}$ to be the \emph{ordered} set of nodes supplying information
to node $i$, i.e, the neighbourhood of $i$.  Without loss of generality,
suppose the elements of $\mathcal{V}_i$ are ordered in the ascending
order. The cardinality of $\mathcal{V}_i$, known as the in-degree of node
$i$, is denoted $p_i$; i.e., $p_i$ is equal to the number of incoming edges
for node $i$. Also, the out-degree of node $i$ (.e., the number of outgoing
edges) is denoted $q_i$.

Without loss of generality the graph $\mathcal{G}$ will be assumed to be
weakly connected, that is, for every two nodes of $\mathcal{G}$ there is an
undirected path between these two nodes. The rationale for this assumption
is based on \cite[Proposition~1]{U6}; according to that proposition
$H_\infty$ consensus optimization problems over disconnected graphs are
reducible to the corresponding problems over individual weakly connected
components.

Let $\mathcal{A}=[\mathbf{a}_{ij}]_{i,j=1}^N$ be the adjacency matrix of the
digraph $\mathcal{G}$, i.e., $\mathbf{a}_{ij}=1$ if $(j,i)\in \mathcal{E}$,
otherwise $\mathbf{a}_{ij}=0$. Also, let $\mathcal{L}$ be the $N\times N$
Laplacian matrix of the graph $\mathcal{G}$,
$\mathcal{L}=\diag[p_1,\ldots,p_N]-\mathcal{A}$.

In the sequel,  a shift permutation operator defined on elements of the set
$\mathcal{V}_i$ will be used:
\begin{equation}
  \label{permu}
  \Pi\{j_1,\ldots,j_{p_i-1},j_{p_i}\}=\{j_{p_i},j_1,\ldots,j_{p_i-1}\}.
\end{equation}
Furthermore, $\Pi^k(\mathcal{V}_i)$ will denote the set obtained from
$\mathcal{V}_i$ using $k$ consecutive shift permutations (\ref{permu}).
In regard to this set, the following notation will be used throughout the
paper unless stated otherwise: for $\nu\in \{1,\ldots, p_i\}$,
$j_\nu$ is the $\nu$-th element in the ordered set
$\mathcal{V}_i$. Conversely,
$\nu_j^{k,i}\in \{1,\ldots, p_i\}$ is the index of element $j$ in the
permutation $\Pi^k(\mathcal{V}_i)$. We will omit the superscript $^{k,i}$
if this does not lead to ambiguity.

\subsection{Distributed estimation with $H_\infty$
  consensus}\label{Estimation}

Consider a plant described  by the equation
\begin{equation}
  \label{eq:plant}
  \dot x=Ax+B_2w(t).
\end{equation}
Here $x\in\mathbf{R}^n$ is the state of the plant, and
$w(t)\in\mathbf{R}^{m_w}$ is a disturbance.
We also assume $w(t)\in L_2[0,\infty)$, so that the $L_2$-integrable solution
of (\ref{eq:plant}) with the initial condition $x(0)=x_0$ exists on any
finite interval $[0,T]$~\cite[p.125]{CP-1977}. Furthermore, it will be
convenient to assume that the plant was at the state $x_0$ for all $t\le 0$.

The distributed filtering problem under consideration is to
estimate the state of the system (\ref{eq:plant})
using a network of filters connected according to the graph
$\mathcal{G}$. Each node takes measurements
\begin{equation}\label{U6.yi}
y_i(t)=C_ix(t)+D_{2i}w(t)+\bar D_{2i}v_i(t);
\end{equation}
$v_i(t)\in\mathbf{R}^{m_v}$ is a measurement disturbance. As seen from
(\ref{U6.yi}), the measurements are assumed to be taken continuously. Although
in practice, measurements are usually taken at discrete time instances,
we assume that the data rate of the sensors is high enough to allow
for the continuous-time interpretation of the measurement signals $y_i$.
This will enable us to focus exclusively on the effects due to sampling and
intermittence of interconnections.

The measurements are processed by a network of observers connected over the
graph $\mathcal{G}$. The key assumption in this paper is to allow the
observers make use of their local
 measurements continuously, however they can only interact with each other
 at discrete time instances $t_k$, $k=0,1\ldots$, with $t_0=0$. For
 simplicity, we assume that this schedule of updates is known to all
 participants in the network, and therefore all nodes
 exchange information at the same time instance $t_k$. However, at every
 time instance $t_k$ only one neighbour in the set $\mathcal{V}_i$ is polled by
 each node $i$, according to the `Round-Robin' rule. Formally, this leads
 us to define the following observer protocol: For $t\in[t_k,t_{k+1})$,
 $k=0,1,\ldots$,
 \begin{eqnarray}
    \dot{\hat x}_i&=&A\hat x_i(t) + L_i(y_i(t)-C_i\hat x_i(t)) \nonumber \\
         &&+K_i\sum_{j\in \Pi^k(\mathcal{V}_i)} H_i(\hat x_j(t_{k-\nu_j^{k,i}+1})-\hat
           x_i(t_{k-\nu_j^{k,i}+1}),
  \label{UP7.C.d}
\end{eqnarray}
where $ \hat{x}_i(t)$ is the estimate of the plant
state $x(t)$ calculated at node $i$,
the matrices $L_i$,
$K_i$ are parameters of the filters to be determined, and $H_i$ is a given
matrix. All observers are initiated with zero initial condition, $\hat
x_i(t)=0$ for all $t\le 0$ and all $i=1,\ldots,N$. In particular, this
ensures that in (\ref{UP7.C.d}), the terms sampled at times
$t_{k-\nu_j^{k,i}+1}<0$ are equal to zero.

From now on, we will omit the time variable when a signal is
considered at time $t$, and will write, for example, $\hat x_i$ for $\hat
x_i(t)$.

The last term in (\ref{UP7.C.d}) reflects the desire of each node observer to
update its estimate of the plant using feedback from the neighbours in
its neighbourhood, according to the consensus estimation
paradigm~\cite{Olfati-Saber-2007,U6}. However, unlike these references,
under the proposed protocol, only one neighbour is polled
at each time $t_k$ to provide a `neighbour feedback', and this sample is
stored and used by the observer until time $t_{k+p_i}$. The feature of the
proposed Round-Robin type protocol is to poll the neighbours one at a time,
in a cyclic manner. Formally, this can be described
by first applying the shift permutation operator $\Pi$ to the neighbourhood
set at every time instance $t_k$, and then selecting the first element from the
resulting permutation $\Pi^k(\mathcal{V}_i)$ for feedback.

Let $e_i=x-\hat x_i$ be the local estimation error at node $i$. This error
satisfies the equation:
 \begin{eqnarray}
    \dot{e}_i&=&(A - L_iC_i)e_i+(B-L_iD_i)\xi_i \nonumber \\
      &+&K_iH_i \sum_{j\in \Pi^k(\mathcal{V}_i)}
      (e_j(t_{k-\nu_j^{k,i}+1})-e_i(t_{k-\nu_j^{k,i}+1})).
\label{e.1}
\end{eqnarray}
Here we used the notation $\xi_i$ to represent the perturbation vector
$[w'~v_i']'$, and the matrices $B$, $D_i$ are defined as follows $B=[B_2~0]$,
$D_i=[D_{2i}~\bar D_{2i}]$. Since the plant was at the
state $x(t)=x_0$ for all $t\le 0$, the initial conditions for (\ref{e.1}) are
$e_i(t)=x_0$ $\forall t\le 0$.

Since the error dynamics (\ref{e.1}) are governed by $L_2$ integrable
disturbance signals $\xi_i$, we can only expect the node observers to converge
in $L_2$ sense. To quantify transient consensus performance of the observer
network (\ref{UP7.C.d}) under disturbances, consider the cost of
disagreement between the observers caused by a particular vector of disturbance
signals $\xi(\cdot)=[\xi_1(\cdot)'~\ldots~\xi_N(\cdot)']'$,
\begin{eqnarray}
  \label{Psi.L2}
  J(\xi)&=&\frac{1}{N} \int_0^\infty \sum_{i=1}^N \sum_{j\in
  \Pi^k(\mathcal{V}_i)}\|\hat x_j(t)-\hat x_i(t)\|^2 dt \nonumber  \\
&=& \frac{1}{N} \int_0^\infty \sum_{i=1}^N \sum_{j\in
  \Pi^k(\mathcal{V}_i)}\|e_j(t)-e_i(t)\|^2dt,
\end{eqnarray}
where $k$ is a time-dependent index, $k=0,1,\ldots$, defined so that for every
$t\in[0,\infty)$, $t_k\le t<t_{k+1}$. The functional (\ref{Psi.L2}) was
originally introduced in~\cite{U6} as a measure of consensus performance of
a corresponding continuous-time observer network. It is worth noting that
for each $t$, $\sum_{j\in \Pi^k(\mathcal{V}_i)}\|\hat x_j(t)-\hat
x_i(t)\|^2$ is independent of the order in which node $i$ polls its
neighbours, so that
\[
\sum_{j\in \Pi^k(\mathcal{V}_i)}\|\hat x_j(t)-\hat
x_i(t)\|^2=\sum_{j\in \mathcal{V}_i}\|\hat x_j(t)-\hat
x_i(t)\|^2.
\]
Therefore, the inner summation in
(\ref{Psi.L2}) can be replaced with summation over the neighbourhood set
$\mathcal{V}_i$. This observation leads to the same expression for $J(\xi)$
as in the case of continuous-time networks~\cite{U6},
\begin{equation}
  \label{Psi.L2.1}
  J(\xi)=\frac{1}{N} \int_0^\infty \sum_{i=1}^N\left
    [(p_i+q_i)\|e_i(s)\|^2-2e_i'\sum_{j\in \mathcal{V}_i}e_j(s) \right] ds.
\end{equation}

The following distributed estimation problem is a version of the
distributed $H_\infty$ consensus-based estimation problem originally
introduced in~\cite{U6,LaU1}, modified to include the Round-Robin type protocol
(\ref{UP7.C.d}).

\begin{definition}\label{Def1}
The distributed estimation problem under consideration is to
determine a collection of observer gains
$L_i$ and interconnection coupling gains
$K_i,~i=1,\ldots,N$, for the filters
(\ref{UP7.C.d}) which ensure that the following conditions are
satisfied:
  \begin{enumerate}[(i)]
  \item
In the absence of uncertainty, the interconnection of unperturbed
systems (\ref{e.1}) must be exponentially stable.

\item
The filter must ensure a specified level of transient consensus performance, as
follows
\begin{eqnarray}\label{objective.i.1}
&&\sup_{x_0, \xi\neq 0}
\frac{J(\xi)}{\|x_0\|^2_P+\frac{1}{N}\|\xi\|_2^2} \le
\gamma^2.
\end{eqnarray}
Here, $\|x_0\|^2_P=x_0'Px_0$, $P=P'>0$ is a matrix to be determined,
and $\gamma>0$ is a given constant.
\end{enumerate}
In~\cite{U6}, the quantity on the left-hand-side of (\ref{objective.i.1})
was referred to as the mean-square $L_2$ disagreement gain of the
distributed observer.
\end{definition}

Note that unlike~\cite{U6,LaU1}, here we aim to achieve internal
stability and $H_\infty$ performance of the observer using a different
communication
protocol, which involves sampling of observer inputs according to the
Round-Robin rule.

\section{The main results}\label{main}

Our approach to solving the problem in Definition~\ref{Def1} will
follow the methodology for the analysis of stability and $L_2$-gain
for networked control systems proposed in~\cite{LFH-2012}. In this
paper, this methodology is further extended to derive
\emph{synthesis conditions} for a network of observers. The
methodology in~\cite{LFH-2012} makes use of the time-delay approach
to sampled-data control started in \cite{Fridman-Aut04}. In
\cite{LFH-2012} the closed-loop system under consideration is
presented as a switched system with multiple and ordered
time-varying delays.

As can be seen from (\ref{e.1}), if the
observer at node $i$ polls a channel at time $t_{k-p_i+1}$, the next time
the same channel will be polled at time $t_{k+1}$. The longest time between
polls of the same channel at node $i$ constitutes the maximum delay in
communication between node $i$ and its neighbours, which will be denoted
$\tau_i$:
\[
\tau_i=\max_{k} (t_{k+1}-t_{k-p_i+1}).
\]
The largest communication delay in the network is then
$\tau=\max_{i}\tau_i$. It is easy to see from these definitions that
$\tau=\max_{k} (t_{k+1}-t_{k-\bar p+1})$, where $\bar p=\max_i p_i$.

Consider the following Lyapunov-Krasovskii candidate for the system
(\ref{e.1}):
\begin{eqnarray}
  V_i(e_i)&=& e_i'Y_i^{-1}e_i+\int_{t-\tau_i}^t\!\! e^{-2\alpha_i(t-s)}
  e_i(s)'S_ie_i(s)ds
   \nonumber \\
   &+&\tau_i\int_{t-\tau_i}^t\!\! e^{-2\alpha_i(t-s)}\dot
   e_i(s)'(\tau_i+s-t)R_i\dot e_i(s)ds,\quad
  \label{Vi}
\end{eqnarray}
where $Y_i=Y_i'>0$,
$R_i=R_i'\ge 0$, $S_i=S_i'\ge 0$ and
$\alpha_i\ge 0$, $i=1,\ldots,N$,
are matrices and constants to be
determined. $V_i(e_i)$ is a standard Lyapunov-Krasovskii functional used in the
literature on  exponential stability of systems with time-varying
delays; e.g., see \cite{LFH-2012}.

Given a matrix $W_i=W_i>0$, define
\[
\mathcal{W}_i(u,z)={\pi^2\over 4}(u-z)'W_i(u-z).
\]

\begin{theorem}\label{dissip-ineq}
Suppose there exist gains $K_i$, $L_i$, matrices $W_i=W_i'>0$, and constants
$\alpha_i>0$, $0<\pi_i<2\alpha_iq_i^{-1}$, $i=1,\ldots,N$, such that the
following vector dissipation inequality holds for all
$i=1,\ldots,N$: For $t\in[t_k,t_{k+1})$,
\begin{eqnarray}
  \lefteqn{\dot V_i(e_i)+2\alpha_iV_i(e_i)-\sum_{j\in \mathcal{V}_i} \pi_j
  V_j(e_j)} && \nonumber \\
&& +\left(\sum_{j:i\in \mathcal{V}_j}\tau_j^2\right)\dot e_i'W_i\dot e_i
 -\sum_{j\in \mathcal{V}_i}\mathcal{W}_j(e_j,e_j(t_{k-\nu_j^{k,i}+1}))
  \nonumber \\
&&  +\frac{1}{\gamma^2}(p_i+q_i)\|e_i\|^2 -\frac{2}{\gamma^2}e_i'\sum_{j\in
    \mathcal{V}_i} e_j - \|\xi_i\|^2\leq 0,
  \label{VLF}
\end{eqnarray}
where $\nu_j^{k,i}$ is the
index of $j$ in the ordered permutation set $\Pi^k(\mathcal{V}_i)$.
Then the system~(\ref{e.1}) satisfies conditions (i) and (ii) in
Definition~\ref{Def1}.
\end{theorem}

The proof of this theorem and other statements are given in the Appendix.

\begin{remark}
Let $\mathcal{V}_i=\{j_1,\ldots, j_{p_i}\}$, and define
\begin{eqnarray*}
\lefteqn{\mathcal{S}_i(e_i,\dot e_i,e_{j_1},\ldots,e_{j_{p_i}},\xi_i)} && \\
&&=\left(\sum_{j:i\in \mathcal{V}_j}\tau_j^2\right)\dot e_i'W_i\dot e_i
 -\sum_{j\in \mathcal{V}_i}\mathcal{W}_j(e_j,e_j(t_{k-\nu_j^{k,i}+1}))
  \nonumber \\
&&  +\frac{1}{\gamma^2}(p_i+q_i)\|e_i\|^2 -\frac{2}{\gamma^2}e_i'\sum_{j\in
    \mathcal{V}_i} e_j - \|\xi_i\|^2.
\end{eqnarray*}
Then, inequality (\ref{VLF}) can be written in  the standard form of a
vector dissipation inequality~\cite{U6},
\begin{eqnarray*}
\dot V_i(e_i)+2\alpha_iV_i(e_i)-\sum_{j\in \mathcal{V}_i} \pi_j
  V_j(e_j) \qquad\qquad \\
\le - \mathcal{S}_i(e_i,\dot e_i,e_{j_1},\ldots,e_{j_{p_i}},\xi_i).
\end{eqnarray*}
This prompts for an interpretation of $V(e)=[V_1(e_1),\ldots, V_N(e_N)]'$ and
$ [\mathcal{S}_1, \ldots, \mathcal{S}_N]'$ as, respectively,
a vector storage function and a vector supply rate for the large scale
system comprised of the error dynamics subsystems
(\ref{e.1})~\cite{HCN-2004,U6}. Strictly
speaking, in our case such an interpretation is somewhat artificial, since
for example, the
derivative signal $\dot e_i$ is not an output of `subsystem' $i$, and is
not used for feedback by any of the neighbours this node. Nonetheless, in
the proof of Theorem~\ref{dissip-ineq} the functions $\mathcal{S}_i$ will
play a role analogous to that played by generalized supply rates
in~\cite{LCD-2004,U6,LaU1}.
\end{remark}

In what follows we derive a sufficient condition for the
dissipation inequality (\ref{VLF}) to hold.
We begin with a technical lemma which essentially restates the
corresponding lemma of~\cite{PKJ-2011} in the form convenient for the
subsequent use in the paper. Consider a vector
$\delta=[\delta_0',\ldots,\delta_{p_i}']'$, $\delta_\nu\in \mathbf{R}^n$.
Also, for given $n\times n$ matrices $R_i=R_i'\ge 0$ and $G_i$, define
\[
\Psi_i=\left[
  \begin{array}{cccc}
    R_i & \frac{1}{2}(G_i+G_i') & \ldots & \frac{1}{2}(G_i+G_i') \\
   \frac{1}{2}(G_i+G_i') & R_i  & \ldots & \frac{1}{2}(G_i+G_i') \\
    \vdots & \vdots & \ddots & \vdots \\
   \frac{1}{2}(G_i+G_i') & \frac{1}{2}(G_i+G_i')  & \ldots & R_i
 \end{array}\right].
\]

\begin{lemma}\label{Lem1}
Suppose the matrices $R_i=R_i'\ge 0$ and $G_i$ are such that
\begin{eqnarray}
\label{Park}
  \left[\begin{array}{cc}R_i & G_i\\ G_i' & R_i
    \end{array}\right]\ge 0.
\end{eqnarray}
Then
\begin{eqnarray*}
 \tau_i \left[
\frac{1}{t-t_k}\delta_0'R_i\delta_0
+ \sum_{\nu=1}^{p_i-1}\frac{1}{t_{k-\nu+1}-t_{k-\nu}}
\delta_\nu'R_i\delta_\nu\right.&&\\
\left.+\frac{1}{t_{k-p_i+1}-t+\tau_i }
\delta_{p_i}'R_i\delta_{p_i}\right]
&\ge& \delta' \Psi_i \delta.
\end{eqnarray*}
\end{lemma}

Let $\mathbf{e}_i=[e_i(t_k)'~\ldots~e_i(t_{k-p_i+2})'~e_i(t_{k-p_i+1})']'$,
$\bar{\mathbf{e}}_i=[e_i'~\mathbf{e}_i'~e_i(t-\tau_i )']'$,
and
$T_i\in R^{(p_i+1)n\times (p_i+2)n}$ be the following matrix
\[
T_i=\left[\begin{array}{rrrrrr} 1 & -1 &  0 & \ldots & 0 & 0 \\
                              0 & 1  & -1 & \ldots & 0 & 0 \\
                              \ldots &   \ldots & \ldots &  \ldots &
                              \ldots & \ldots \\
                              0 & 0  & 0  & \ldots & 1 & -1
                            \end{array}\right]\otimes I.
\]
Define $\bar\Psi_i=e^{-2\alpha_i\tau_i } T_i'\Psi_i T_i$ and partition this
matrix in accordance with the partition of
$\bar{\mathbf{e}}_i$:
\[
\bar\Psi_i=e^{-2\alpha_i\tau_i } T_i'\Psi_i T_i=\left[\begin{array}{ccc}
\bar\Psi_{i,11} & \bar\Psi_{i,12} & \bar\Psi_{i,13} \\
\bar\Psi_{i,12}' & \bar\Psi_{i,22} & \bar\Psi_{i,23} \\
\bar\Psi_{i,13}' & \bar\Psi_{i,23}' & \bar\Psi_{i,33} \\
\end{array}\right].
\]
Also, let us introduce the correspondingly partitioned matrix
\begin{eqnarray}
&&
\tilde\Psi_i=\left[\begin{array}{ccc}
\tilde\Psi_{i,11} & \tilde\Psi_{i,12} & \tilde\Psi_{i,13} \\
\tilde\Psi_{i,12}' & \tilde\Psi_{i,22} & \tilde\Psi_{i,23} \\
\tilde\Psi_{i,13}' & \tilde\Psi_{i,23}' & \tilde\Psi_{i,33} \\
\end{array}\right], \label{tildePsi}
\end{eqnarray}
where we let
$\tilde\Psi_{i,11}= \bar\Psi_{i,11}-2\alpha_iY_i^{-1}-S_i$,
$\tilde\Psi_{i,33}= \bar\Psi_{i,33}+e^{-2\alpha_i\tau_i}S_i$,
and $\tilde\Psi_{i,\mu\nu}=\bar\Psi_{i,\mu\nu}$ for all other
  elements of $\tilde\Psi_i$. Then the following statement holds.

\begin{lemma}\label{dotV.lemma}
Under the conditions of Lemma~\ref{Lem1},
\begin{eqnarray}
  \dot V_i &\le&
-2\alpha_i V_i(e_i) + 2e_i'Y_i^{-1}\dot e_i \nonumber \\
&& +\tau_i ^2\dot
   e_i(t)'R_i\dot e_i(t) - \bar{\mathbf{e}}_i'\tilde \Psi_i\bar{\mathbf{e}}_i.
\label{dotVi.2}
\end{eqnarray}
\end{lemma}

Furthermore, since $R_j,S_j\ge 0$, for every $j\in\mathcal{V}_i$, we have
\begin{eqnarray}
 -\pi_j V_j(e_j) \le -\pi_j  e_j'Y_j^{-1}e_j. \label{pi_jV_j}
\end{eqnarray}
This leads to the following statement.

\begin{lemma}\label{pi_jV_j.lemma}
\begin{eqnarray}
-\sum_{j\in\mathcal{V}_i} \pi_j V_j(e_j)-
\sum_{j\in\mathcal{V}_i}\mathcal{W}_j(e_j,e_j(t_{k-\nu_j^{k,i}+1}))
\nonumber
\\
\le  - \left[\begin{array}{c} \mathbf{e}_{i,t} \\
\mathbf{e}_{i,s}
\end{array} \right]'
\left[\begin{array}{cc}
\bar \Phi_{i,11} & \bar \Phi_{i,12} \\
\bar \Phi_{i,21} & \bar \Phi_{i,22}
\end{array} \right]
\left[\begin{array}{c}
\mathbf{e}_{i,t} \\  \mathbf{e}_{i,s}
\end{array} \right],
\label{pi_jV_j.1}
\end{eqnarray}
where
\begin{eqnarray*}
&&\bar \Phi_{i,11}= \left[\begin{array}{ccc}
\pi_{j_1}Y_{j_1}^{-1}+{\pi^2\over 4}W_{j_1} & \ldots & 0 \\
\vdots & \ddots & \vdots \\
0 & \ldots & \pi_{j_{p_i}}Y_{j_{p_i}}^{-1}+{\pi^2\over
4}W_{j_{p_i}}
\end{array} \right], \\
&&\bar \Phi_{i,22}=
\left[\begin{array}{cccc}
{\pi^2\over 4}W_{j_1}  & 0 & \ldots & 0 \\
0 & {\pi^2\over 4}W_{j_2}  & \ldots & 0 \\
\vdots & \vdots & \ddots & \vdots \\
0 & 0 & \ldots & {\pi^2\over 4}W_{j_{p_i}}
\end{array} \right], \\
&&\bar \Phi_{i,12}=\bar \Phi_{i,21}=-\bar \Phi_{i,22}.
\end{eqnarray*}
\end{lemma}

Next, we apply the descriptor method~\cite{Fridman-2001} in order to derive
LMIs for the design of observers' gains. For $t\in [t_k,t_{k+1})$,
consider the neighbourhood set  $\mathcal{V}_i$ and its
corresponding permutation $\Pi^k(\mathcal{V}_i)$. Recall that for
every $j\in\mathcal{V}_i$, $\nu_j^{k,i}\in\{1,\ldots,p_i\}$ is the
index of node $j$ in
the ordered set $\Pi^k(\mathcal{V}_i)$.
According to this notation, on the interval $[t_k,t_{k+1})$
the observer at node $i$ utilizes the sample $\hat x_j(t_{k-\nu_j^{k,i}+1})$,
and the corresponding error equation is driven by $e_j(t_{k-\nu_j^{k,i}+1})$.
Let us define vectors
\begin{eqnarray*}
&&\mathbf{e}_{i,t}=[e_{j_1}(t)'~\ldots~
e_{j_{p_i-1}}(t)'~
e_{j_{p_i}}(t)']', \\
&&\mathbf{e}_{i,s}=[e_{j_1}(t_{k-\nu_{j_1}^{k,i}+1})'~\ldots~
e_{j_{p_i}}(t_{k-\nu_{j_{p_i}}^{k,i}+1})']',
\end{eqnarray*}
which consist of the current and sampled error interconnection
inputs, respectively, ordered in accordance with ordering of the set
$\mathcal{V}_i$.
Note that for arbitrary compatible matrices $X_i$, $Z_i$ and $Q_i$,
\begin{eqnarray}
\lefteqn{(X_ie_i+Z_i\dot e_i
+(\mathbf{1}_{p_i}'\otimes Q_i) \mathbf{e}_{i,s} )'} &&
\nonumber \\
&&
\times \left((A - L_iC_i)e_i+(\mathbf{1}_{p_i}'\otimes K_iH_i)
  \mathbf{e}_{i,s}\right. \nonumber \\
&& \left.
- (\mathbf{1}_{p_i}'\otimes K_iH_i) \mathbf{e}_i +(B-L_iD_i)\xi_i
-\dot{e}_i\right)=0.
\label{descr}
\end{eqnarray}
From (\ref{dotVi.2}) and (\ref{descr}) it follows that
\begin{eqnarray*}
 \dot V_i&+&2\alpha_i V_i(e_i)
\le
 2e_i'Y_i^{-1}\dot e_i
+\tau_i ^2\dot
   e_i' R_i\dot e_i
- \bar{\mathbf{e}}_i'\tilde \Psi_i\bar{\mathbf{e}}_i  \\
&& + (X_ie_i+Z_i\dot e_i
+(\mathbf{1}_{p_i}'\otimes Q_i) \mathbf{e}_{i,s} )'
\nonumber \\
&&
\times \left((A - L_iC_i)e_i+(\mathbf{1}_{p_i}'\otimes K_iH_i)
  \mathbf{e}_{i,s}\right. \nonumber \\
&& \left.
- (\mathbf{1}_{p_i}'\otimes K_iH_i) \mathbf{e}_i +(B-L_iD_i)\xi_i
-\dot{e}_i\right).
\end{eqnarray*}
Along with condition (\ref{pi_jV_j.1}) established in
Lemma~\ref{pi_jV_j.lemma}, this leads to the conclusion that
\begin{eqnarray}
\lefteqn{ \dot V_i(e_i)+2\alpha_iV_i-\sum_{j\in\mathcal{V}_i} \pi_j
  V_j(e_j)} & \nonumber \\
&&
  +\left(\sum_{j:~i\in \mathcal{V}_j}\tau_j^2\right)\dot
   e_i'W_i\dot e_i-
\sum_{j\in \mathcal{V}_i}
\mathcal{W}_j(e_j,e_j(t_{k-\nu_j^{k,i}+1}))
\nonumber \\
&&
  +\frac{1}{\gamma^2}(p_i+q_i)\|e_i\|^2 - \frac{2}{\gamma^2}e_i'\sum_{j\in
    \mathcal{V}_i} e_j - \|\xi_i\|^2   \nonumber \\
&&
\le
\eta_i' \Xi_i\eta_i.
\label{VLF1}
\end{eqnarray}
In the above inequality, $\eta_i$ is the vector
$\eta_i=[\dot e_i'~ e_i'~ \mathbf{e}_i'~
e_i(t-\tau_i)'~\mathbf{e}_{i,t}'~ \mathbf{e}_{i,s}'~\xi']'$, and $\Xi_i$ is
the matrix partitioned as follows
\begin{eqnarray}
\label{Xi_i}
\Xi_i=
\left[
\begin{array}{ccccccc}
\Xi_{aa} & \Xi_{ab} & \Xi_{ac} & 0 & 0 & \Xi_{af} & \Xi_{ag}\\
\star & \Xi_{bb} & \Xi_{bc} & -\tilde\Psi_{i,13} & \Xi_{be} & \Xi_{bf} & \Xi_{bg}\\
\star & \star & -\tilde\Psi_{i,22} & -\tilde\Psi_{i,23} & 0 & \Xi_{cf} & 0 \\
\star & \star & \star & -\tilde\Psi_{i,33} & 0 & 0 & 0 \\
\star & \star & \star & \star & -\bar\Phi_{i,11} &  -\bar\Phi_{i,12} & 0 \\
\star & \star & \star & \star & \star & \Xi_{ff} & \Xi_{fg}\\
\star & \star & \star & \star & \star & \star & -I
\end{array}\right],
\end{eqnarray}
\begin{eqnarray*}
\Xi_{aa}&=&\tau_i^2R_i+\left(\sum_{j:~i\in\mathcal{V}_j}\tau_j^2\right) W_i
-Z_i-Z_i', \\
\Xi_{ab}&=&Y_i^{-1}-X_i+Z_i'(A-L_iC_i), \\
\Xi_{ac}&=&-Z_i'(\mathbf{1}_{p_i}'\otimes K_iH_i),\quad
\Xi_{af}=\mathbf{1}_{p_i}'\otimes (-Q_i+Z_i'K_iH_i),\\
\Xi_{ag}&=&Z_i'(B-L_iD_i),\\
\Xi_{bb}&=&
\frac{(p_i+q_i)}{\gamma^2}I-\tilde \Psi_{i,11} +X_i'(A-L_iC_i)+(A-L_iC_i)'X_i,
\\
\Xi_{bc}&=&-\tilde\Psi_{i,12}-(\mathbf{1}_{p_i}'\otimes X_i'K_iH_i), \quad
\Xi_{be}=-\frac{1}{\gamma^2}(\mathbf{1}_{p_i}'\otimes I), \\
\Xi_{bf}&=&\mathbf{1}_{p_i}'\otimes(X_i'K_iH_i+(A-L_iC_i)'Q_i),\\
\Xi_{bg}&=&X_i'(B-L_iD_i),\quad
\Xi_{cf}=-\mathbf{1}_{p_i}\mathbf{1}_{p_i}'\otimes (H_i'K_i'Q_i),\\
\Xi_{ff}&=&\mathbf{1}_{p_i}\mathbf{1}_{p_i}'\otimes (Q_i'K_iH_i+H_i'K_i'Q_i)
-\bar \Phi_{i,22}, \\
\Xi_{fg}&=&\mathbf{1}_{p_i}\otimes Q_i'(B-L_iD_i).
\end{eqnarray*}

It is worth noting that the matrix $\Xi_i$ does not depend on $k$. Hence
the dissipation inequality follows from the condition $\Xi_i<0$ at any
time $t$. By combining this conclusion with Theorem~\ref{dissip-ineq}, we
arrive at the following statement.

\begin{theorem}\label{analysis}
Suppose there exist matrices $Y_i=Y_i'>0$, $X_i$, $Z_i$, $Q_i$,
$W_i=W_i'\ge 0$, $S_i=S_i'\ge 0$, $R_i=R_i'\ge 0$, $G_i$, constants
$\alpha_i>0$, $0\le \pi_i<2\alpha_i q_i^{-1}$, and gain matrices $K_i,L_i$,
$i=1,\ldots, N$, which satisfy  the LMI (\ref{Park}) and
\begin{eqnarray}
\label{LMI.analysis}
\Xi_i<0.
\end{eqnarray}
Then the corresponding observer network (\ref{UP7.C.d}) solves the problem posed in
Definition~\ref{Def1}. The
matrix $P$ in condition (\ref{objective.i.1})
corresponding to this solution is
$P=
\frac{1}{N}\sum_{i=1}^N(Y_i^{-1}
+S_i\frac{1-e^{-2\alpha_i\tau_i}}{2\alpha_i})$.
\end{theorem}

Theorem~\ref{analysis} serves as the basis for derivation of
the main result of this paper, given below in Theorem~\ref{T1}, which is
a sufficient
condition for synthesis of distributed observer networks of the form
(\ref{UP7.C.d}). Consider the following matrix
\begin{eqnarray}
\label{barXi_i}
\bar\Xi_i=
\left[
\begin{array}{ccccccc}
\bar\Xi_{aa} & \bar\Xi_{ab} & \bar\Xi_{ac} & 0 & 0 & \bar\Xi_{af} & \bar\Xi_{ag}\\
\star & \bar\Xi_{bb} & \bar\Xi_{bc} & -\tilde\Psi_{i,13} & \bar\Xi_{be} & \bar\Xi_{bf} & \bar\Xi_{bg}\\
\star & \star & -\tilde\Psi_{i,22} & -\tilde\Psi_{i,23} & 0 & \bar\Xi_{cf} & 0 \\
\star & \star & \star & -\tilde\Psi_{i,33} & 0 & 0 & 0 \\
\star & \star & \star & \star & -\bar\Phi_{i,11} &  -\bar\Phi_{i,12} & 0 \\
\star & \star & \star & \star & \star & \bar\Xi_{ff} & \bar\Xi_{fg}\\
\star & \star & \star & \star & \star & \star & -I
\end{array}\right],
\end{eqnarray}
\begin{eqnarray*}
\bar\Xi_{aa}&=&\tau_i^2R_i+\left(\sum_{j:~i\in\mathcal{V}_j}\tau_j^2\right) W_i
-\epsilon_i X_i-\epsilon_i X_i', \\
\bar\Xi_{ab}&=&Y_i^{-1}-X_i+\epsilon_i (X_i'A-U_iC_i), \\
\bar\Xi_{ac}&=&-\epsilon_i (\mathbf{1}_{p_i}'\otimes F_iH_i),\\
\bar\Xi_{af}&=&\mathbf{1}_{p_i}'\otimes (-\bar\epsilon_i X_i+\epsilon_i
F_iH_i),\quad
\bar\Xi_{ag}=\epsilon_i (X_i'B-U_iD_i),\\
\bar\Xi_{bb}&=&
\frac{(p_i+q_i)}{\gamma^2}I-\tilde \Psi_{i,11} +X_i'A-U_iC_i+A'X_i-C_i'U_i',
\\
\bar\Xi_{bc}&=&-\tilde\Psi_{i,12}-\mathbf{1}_{p_i}'\otimes (F_iH_i), \quad
\bar\Xi_{be}=-\frac{1}{\gamma^2}(\mathbf{1}_{p_i}'\otimes I), \\
\bar\Xi_{bf}&=&\mathbf{1}_{p_i}'
      \otimes(F_iH_i+\bar\epsilon_i AX_i-\bar\epsilon_i C'_iU_i'),\\
\bar\Xi_{bg}&=&X_i'B-U_iD_i,\quad
\bar\Xi_{cf}=-\mathbf{1}_{p_i}\mathbf{1}_{p_i}'\otimes (\bar\epsilon_i H_i'F_i'),\\
\bar\Xi_{ff}&=&\bar\epsilon_i\mathbf{1}_{p_i}\mathbf{1}_{p_i}'
           \otimes (F_iH_i+H_i'F_i')
-\bar \Phi_{i,22}, \\
\bar\Xi_{fg}&=&\bar\epsilon_i\mathbf{1}_{p_i}\otimes (X_i'B-U_iD_i).
\end{eqnarray*}

\begin{theorem}\label{T1}
Suppose there exists matrices $Y_i=Y_i'>0$, $X_i$, $\det X_i\neq 0$,
$F_i$, $U_i$, $S_i=S_i'\ge 0$, $R_i=R_i'\ge 0$, $W_i=W_i'\ge 0$, $G_i$, and
constants $\alpha_i>0$, $0\ge \pi_i<2\alpha_iq_i^{-1}$, $\epsilon_i>0$,
$\bar\epsilon_i>0$, $i=1,\ldots, N$,
which satisfy the LMI (\ref{Park}) and
\begin{eqnarray}
\label{LMI.synthesis}
\bar \Xi_i<0.
\end{eqnarray}
Then the network of observers (\ref{UP7.C.d}) with
\begin{eqnarray}
K_i=(X_i')^{-1}F_i, \quad L_i=(X_i')^{-1}U_i,
\label{gains}
\end{eqnarray}
solves the distributed estimation problem posed in Definition~\ref{Def1}.
The matrix $P$ in condition (\ref{objective.i.1}) corresponding to this
solution is
$P=
\frac{1}{N}\sum_{i=1}^N(Y_i^{-1}+S_i\frac{1-e^{-2\alpha_i\tau_i}}{2\alpha_i})$.
\end{theorem}

\emph{Proof: }  Similar to \cite{SFS-2007},  we observe that LMI
(\ref{LMI.analysis})  follows from (\ref{LMI.synthesis}), when we let
$Z_i=\epsilon_i X_i$, $Q_i=\bar\epsilon_i X_i$, and take $K_i,L_i$ to be
matrices defined in (\ref{gains}). Then the claim of the theorem follows from
Theorem~\ref{analysis}. \hfill$\Box$

\begin{remark}
The proposed LMI conditions involve `free' variables $X_i$, $Z_i$
and $Q_i$. These variables are to reduce the conservatism of the
proposed LMI conditions. At the same time they add to the number of
unknowns to be used by the LMI solver. In a high-dimensional problem where this
causes an excessive computational burden, additional constraints on these
variables can be introduced to reduce the number of variables used by the
solver, at the expense of a more conservative design; e.g., $X_i$ can be
assumed to be diagonal.    
\end{remark}

\section{Example}\label{example}

Consider a plant of the form (\ref{eq:plant}), with
$A=\left[\begin{smallmatrix}
   -3.2 &  10    &  0\\
    1   &  -1    &  1 \\
    0   & -14.87 &  0
  \end{smallmatrix}
\right]$, \quad
$B_2=\left[\begin{smallmatrix}
    -0.1246 \\
   -0.4461 \\
    0.3350
  \end{smallmatrix}
\right]$.
This plant was used in the example in~\cite{U7}.
The nominal part of the plant describes one of the regimes of the so-called
Chua electronic circuit.

\begin{figure}
  \centering
  \psfrag{k0}{$k=0,2,\ldots$}
  \psfrag{k1}{$k=1,3,\ldots$}
  \psfrag{i}{2}
  \psfrag{j2}{1}
  \psfrag{j3}{3}
  \includegraphics[width=6cm]{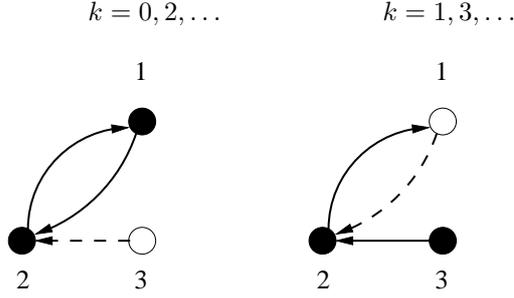}
  \caption{An example 3-node network. The filled circles and
    solid lines represent nodes and links which are `active' during the
    time interval
   $[t_k,t_{k+1})$, when $k$ takes one of the values shown above the figure.}
  \label{fig:example}
\end{figure}

To estimate this plant, we will use the 3-node
observer network shown in Figure~\ref{fig:example}.

The measurement matrices are
\begin{eqnarray*}
&&C_1=[ 0.0032~-0.0047~0.0010], \\
&&C_2=[-0.8986~0.1312~-1.9703], \\
&&C_3=[1~0~0], \quad \mbox{and}\quad D_{2i}=0, \quad \bar D_{2i}=0.025.
\end{eqnarray*}
With these
parameters, the pairs $(A,C_1)$ and $(A,C_2)$ are not detectable, while
$(A,C_3)$ is observable. Since observer at node 3 does not receive
information from other observers, it acts as a conventional continuous-time
$H_\infty$
filter, while the observers at nodes 1 and 2 utilize sampled data inputs they
receive from their neighbours. This allows them to overcome difficulties due to
unstable unobservable modes of $A$.

For simplicity we assume a constant sampling
period of $\Delta$, so that $t_k=k\Delta$. Then
$\tau_1=\Delta$, $\tau_2=2\Delta$ and $\tau_3= 0$.
We now apply Theorem~\ref{T1} to compute observer and interconnection gains
for this system. To this end, we solved the LMIs (\ref{LMI.synthesis})
numerically, with $\alpha_i=0.1$,
$\pi_i=\frac{2\alpha_i}{1+q_i}$, and $\bar\epsilon_i=0$;
that is, $Q_i=0$ in this example. In fact, instead of solving the
feasibility problem, we solved the optimization problem in which we sought
to minimize $\gamma^2$ subject to the LMI constraints (\ref{Park}) and
(\ref{LMI.synthesis}).

First, we compared the performance of our method with the performance guaranteed
for estimators employing continuous-time interconnections by the method in
\cite{LaU1}. To this end, we set the sampling rate to a high value by
letting $\Delta=0.0001$. With $\epsilon_i=0.01$, we obtained the suboptimal
$\gamma^2$ to be equal
$0.2274$, which is approximately equal to the level
of $H_\infty$ disagreement guaranteed for the comparison distributed
estimator of \cite{LaU1}, $\gamma^2=0.2299$. A slight discrepancy
between the two values is likely due to numerical errors and/or
conservative selection of parameters. Remarkably, both algorithms assign a
high gain to the $H_\infty$ filter at node 3 ($L_3=10^3\times
[0.2385~0.4724~3.9685]'$ using Theorem~\ref{T1} versus $L_3=10^3\times
[0.0819~0.1707~1.5540]'$ using the method from \cite{LaU1}).

Next we set the sampling rate to a larger value. After some experimenting with
the tuning parameters $\epsilon_i$, we chose $\epsilon_i=0.1$. With $\Delta=0.1$, Theorem~\ref{T1} was found to guarantee the level
of $H_\infty$ disagreement $\gamma^2=0.5537$, and the gain $L_3$ reduced
substantially, to the value $L_3 =[17.9083~13.1006~-19.6797]'$. This gain
is comparable with that obtained for the estimator of~\cite{LaU1} with this
value of $\gamma^2$.
For $\Delta=0.2$, the guaranteed level
of $H_\infty$ disagreement increased substantially, to
the value of $\gamma^2=39.6506$.
Further increasing the sampling period to $\Delta=0.22$ resulted in a
prohibitively large $\gamma^2=896.9248$.

\section{Conclusions}\label{Concl}

The paper has presented a sufficient LMI condition for the design of a
Round-Robin type
interconnection protocol for networks of distributed observers. We have shown
that the proposed protocol allows one to use sampled-data communications
between the observers in the network, and does not require a combinatorial
gain scheduling. As a result, the node observers are shown to be capable of
achieving the $H_\infty$ consensus objective introduced
in~\cite{U6,LaU1,U7}. As our example demonstrates, the proposed Round-Robin
protocol achieves this  objective at the expense of moderately deteriorated
performance.

\section{Appendix}

\subsection{Proof of Theorem~\ref{dissip-ineq}}

Define the vector function
$V(e)=[V_1(e_1),\ldots V_N(e_N)]'$ and the
matrix $M=[M_{ij}]$, $M_{ii}=-2\alpha_i$, $M_{ij}=\pi_j$ if $j\in
\mathcal{V}_i$, and $M_{ij}=0$ if $j\not\in
\mathcal{V}_i$ and $j\neq i$.
It follows from (\ref{VLF}) that for $t\in[t_k,t_{k+1})$,
\begin{eqnarray}
  \lefteqn{\mathbf{1}_N'(\dot V -MV)+\frac{1}{\gamma^2}\sum_{i=1}^N\sum_{j\in \mathcal{V}_i}\|e_i-e_j\|^2 } && \nonumber \\
&& \le
\sum_{i=1}^N \|\xi_i\|^2 -\sum_{i=1}^N\left(\sum_{j:i\in \mathcal{V}_j}\tau_j^2\right)\dot
e_i'W_i\dot e_i  \nonumber \\
&&
 +\sum_{i=1}^N\sum_{j\in \mathcal{V}_i}
\mathcal{W}_j(e_j,e_j(t_{k-\nu_j^{k,i}+1})).
\label{VLF.sum}
\end{eqnarray}
By changing the order of summation in the second term, we further obtain
\begin{eqnarray*}
\sum_{i=1}^N\left(\sum_{j:i\in \mathcal{V}_j}\tau_j^2\right)\dot e_i'W_i\dot e_i
&=&\sum_{i=1}^N \sum_{j\in\mathcal{V}_i}\tau_i^2 \dot e_j'W_j\dot e_j.
\end{eqnarray*}
Hence, on the time interval $t\in[t_k,t_{k+1})$,
\begin{eqnarray}
\mathbf{1}_N'(\dot V -MV)+\frac{1}{\gamma^2}\sum_{i=1}^N\sum_{j\in \mathcal{V}_i}\|e_i-e_j\|^2 \le
\sum_{i=1}^N \Bigg[\|\xi_i\|^2 \nonumber \\
 - \sum_{j\in\mathcal{V}_i}\Big[
  \tau_i^2 \dot e_j'W_j\dot e_j
 -\mathcal{W}_j(e_j,e_j(t_{k-\nu_j^{k,i}+1}))\Big]\Bigg].
\label{VLF.sum.1}
\end{eqnarray}

Let $T>0$ be a time instant, $T\in[t_{\bar k},t_{\bar k+1})$. Let us fix $i$ and
$j\in\mathcal{V}_i$, and consider the partition of the interval $[0,T]$
into subintervals
\begin{eqnarray}
[0,T]&=&[0,t_{\bar k-\bar dp_i-\bar\nu+1})\cup
[t_{\bar k-\bar\nu+1},T) \nonumber \\
&&\cup\left(\bigcup_{d=1}^{\bar d}
[t_{\bar k-dp_i-\bar\nu+1},t_{\bar k-(d-1)p_i-\bar\nu+1})\right),
\label{partition}
\end{eqnarray}
where $\bar\nu=\nu_j^{\bar k,i}$ is the index of $j$ in the permutation
$\Pi^{\bar k}(\mathcal{V}_i)$, and $\bar d$ is the largest integer
number such that $\bar d\le
\frac{\bar k-\bar\nu+1}{p_i}$. Note that $0\le \bar k-\bar
dp_i-\bar\nu+1$, and $t_{\bar k-\bar\nu+1}\le T<
t_{\bar k-\bar\nu+p_i+1}$. The significance of this partition is that on
each interval $[t_{\bar k-dp_i-\bar\nu+1},t_{\bar k-(d-1)p_i-\bar\nu+1})$,
the observer at node $i$ makes use of the sample $\hat
x_j(t_{\bar k-dp_i-\bar\nu+1})$. Therefore the input $e_j$ into the error
dynamics equation (\ref{e.1}) at node $i$ holds the constant value
$e_j(t_{\bar k-dp_i-\bar \nu+1})$ over this
interval of time, with $e_j(t_l)=x_0=\mathrm{const}$ for $0\le t_l< \bar k-\bar
dp_i-\bar\nu+1$. That is, for
$t\in[t_{\bar k-dp_i-\bar\nu+1},t_{\bar k-(d-1)p_i-\bar\nu+1})$,
\[
\mathcal{W}_j(e_j,e_j(t_{k-\nu_j^{k,i}+1}))
=
\mathcal{W}_j(e_j,e_j(t_{\bar k-dp_i-\bar\nu+1}))
\]
where $k=k(t)$ is determined from the condition $t_k\le t<t_{k+1}$, and
$\nu_j^{k,i}=\nu_j^{k(t),i}$ is determined accordingly, as an index of $j$
in the permutation $\Pi^{k(t)}(\mathcal{V}_i)$.

It follows from the above discussion that
\begin{eqnarray}
\lefteqn{\int_0^T \left(\tau_i^2 \dot e_j'W_j\dot e_j -
\mathcal{W}_j(e_j,e_j(t_{k-\nu_j^{k,i}+1}))\right) dt} && \nonumber \\
&&=
\int_0^{t_{\bar k-\bar d p_i-\bar \nu+1}}
\Big(\tau_i^2 \dot  e_j'(t)W_j\dot e_j(t)
-\mathcal{W}_j(e_j,e_j(0))\Big)dt
\nonumber \\
&& +
\sum_{d=1}^{\bar d} \int_{t_{\bar k-dp_i-\bar \nu+1}}^{t_{\bar
    k-(d-1)p_i-\bar \nu+1}} \Big(\tau_i^2 \dot e_j'W_j\dot e_j \nonumber \\
&&\qquad\qquad\qquad\qquad -\mathcal{W}_j(e_j,e_j(t_{\bar
    k-dp_i-\bar \nu+1})) \Big)dt \nonumber \\
&& +
\int_{t_{\bar k-\bar \nu+1}}^T \!\!\!\!
\Big(\tau_i^2
\dot  e_j'(t)W_j\dot e_j(t)
-\mathcal{W}_j(e_j,e_j(t_{\bar k-\bar \nu+1}))\Big)dt.\quad
\label{sum}
\end{eqnarray}

Using the Wirtinger's inequality~\cite{LSF-2010}, it follows that
\begin{eqnarray}
\label{IMA}
\lefteqn{
\int\limits_{t_{\bar k-dp_i-\bar \nu+1}}^{t_{\bar k-(d-1)p_i-\bar \nu+1}}
\!\!\!
\Big(\tau_i^2\dot  e_j'(t)W_j\dot e_j(t)-\mathcal{W}_j(e_j,e_j(t_{\bar
    k-dp_i-\bar \nu+1})) \Big)dt} && \nonumber \\
&& \ge
(t_{\bar k-(d-1)p_i-\bar \nu+1}-t_{\bar k-dp_i-\bar \nu+1})^2 \!\!\! \int\limits_{t_{\bar k-dp_i-\bar \nu+1}}^{t_{\bar k-(d-1)p_i-\bar \nu+1}}\!\!\!
\Big(\dot  e_j'(t)W_j\dot e_j(t)-\mathcal{W}_j(e_j,e_j(t_{\bar
    k-dp_i-\bar \nu+1})) \Big)dt \nonumber \\
&& \ge 0.
\end{eqnarray}
Similarly,
\begin{eqnarray}
\label{IMA.0}
\int_0^{t_{\bar k-\bar d p_i-\bar \nu+1}}
\Big(\tau_i^2\dot  e_j'(t)W_j\dot e_j(t)
- \mathcal{W}_j(e_j,e_j(0)) \Big)dt \ge 0, \nonumber \\
\int_{t_{\bar k-\bar \nu+1}}^T
\Big(\tau_i^2\dot  e_j'(t)W_j\dot e_j(t)
- \mathcal{W}_j(e_j,e_j(t_{\bar k-\bar \nu+1})) \Big)dt
 \ge 0. \nonumber \\
\end{eqnarray}
Therefore, we conclude from (\ref{sum}), (\ref{IMA}) and (\ref{IMA.0}) that
\begin{eqnarray*}
\int_0^T \left(\tau_i^2 \dot e_j'W_j\dot e_j
- \mathcal{W}_j(e_j,e_j(t_{k-\nu_j^{k,i}+1}))\right) dt \ge 0.
\end{eqnarray*}
Hence, it follows from (\ref{VLF.sum.1}) that
\begin{eqnarray}
 \int_0^T\left(\mathbf{1}_N'(\dot V -MV)\right)dt +
\frac{1}{\gamma^2}\sum_{i=1}^N\sum_{j\in
  \mathcal{V}_i}\int_0^T\|e_i-e_j\|^2dt \nonumber \\
\hspace{-1cm} \le \sum_{i=1}^N \int_0^T \|\xi_i\|^2 dt.\hspace{1cm}
\label{VLF.sum.2}
\end{eqnarray}
The statement of the theorem then follows from (\ref{VLF.sum.2}) using
the same argument as that used in the proof of Theorem~1 in
\cite{U6}.

\subsection{Proof of Lemma~\ref{dotV.lemma}}
Consider $\dot V_i$:
\begin{eqnarray*}
  \dot V_i &=& 2e_i'Y_i^{-1}\dot e_i+e_i'S_ie_i-e^{-2\alpha_i\tau_i}e_i'(t-\tau_i)S_ie_i(t-\tau_i)
\\
&&+\tau_i \int_{t-\tau_i }^t\left[ \dot
   e_i(t)'R_i\dot e_i(t) - e^{2\alpha_i(t-s)} \dot
   e_i(s)'R_i\dot e_i(s)\right]ds \\
&&
-2\alpha_i\int_{t-\tau_i}^t e^{-2\alpha_i(t-s)}
  e_i(s)'S_ie_i(s)ds \\
&&
- 2\alpha_i \tau_i \int_{t-\tau_i }^t e^{-2\alpha_i(t-s)} \dot e_i(s)'(\tau +s-t)R_i
\dot e_i(s)ds.
\end{eqnarray*}

Since $e^{2\alpha_i(t-s)}\ge e^{-2\alpha_i\tau_i }$ for $s\in[t-\tau_i,t]$,
then
\begin{eqnarray*}
  \dot V_i &\le & 
-2\alpha_i V_i(e_i) + 2e_i'Y_i^{-1}\dot e_i
  +e_i'(2\alpha_iY_i^{-1}+S_i)e_i  \\
&&+\tau_i^2\dot e_i(t)'R_i  \dot e_i(t)-e^{-2\alpha_i\tau_i}e_i'(t-\tau_i)S_ie_i(t-\tau_i) \\
 && - \tau_i e^{-2\alpha_i\tau_i }\left[ \int_{t_k}^t \dot
e_i(s)'R_i\dot e_i(s) ds \right. \\
&&+ \sum_{\nu=1}^{p_i-1}
\int_{t_{k-\nu}}^{t_{k-\nu+1}} \dot e_i(s)'R_i\dot e_i(s) ds \\
&&\left.+\int_{t-\tau_i }^{t_{k-p_i+1}} \dot e_i(s)'R_i\dot e_i(s) ds
\right].
\end{eqnarray*}

By Jensen's inequality,
\begin{eqnarray*}
  \dot V_i &\le &
-2\alpha_i V_i(e_i) + 2e_i'Y_i^{-1}\dot e_i
  +e_i'(2\alpha_iY_i^{-1}+S_i)e_i \\
&&\hspace{-.5cm}+\tau_i^2\dot
   e_i(t)' R_i\dot e_i(t) -e^{-2\alpha_i\tau_i}e_i'(t-\tau_i)S_ie_i(t-\tau_i)\\
&& \hspace{-.5cm}- \tau_i e^{-2\alpha_i\tau_i }\left[
\frac{1}{t-t_k}(e_i-e_i(t-t_k))'R_i (e_i-e_i(t-t_k))\right. \\
&&
\hspace{-.5cm}+ \sum_{\nu=1}^{p_i-1}\frac{1}{t_{k-\nu+1}-t_{k-\nu}}
(e_i(t_{k-\nu+1})-e(t_{k-\nu}))' \\
&& \qquad\qquad \times
R_i(e_i(t_{k-\nu+1})-e(t_{k-\nu})) \\
&&\hspace{-.5cm}
+\frac{1}{t_{k-p_i+1}-t+\tau_i }
(e_i(t_{k-p_i+1})-e(t-\tau_i ))' \\
&&\qquad\qquad \times R_i(e_i(t_{k-p_i+1})-e(t-\tau_i ))\bigg].
\end{eqnarray*}
Let $\delta=[\delta_0',\ldots,\delta_{p_i}']'$, where
\begin{eqnarray*}
&& \delta_\nu=e_i(t_{k-\nu+1})-e(t_{k-\nu}), \quad \nu=1,\ldots,p_i-1, \\
&& \delta_0=e_i(t)-e(t_k), \quad\delta_{p_i}=e_i(t_{k-p_i+1})-e(t-\tau_i ).
\end{eqnarray*}
Then, $\delta= T_i\bar{\mathbf{e}}_i$. Also, $\delta'\Psi_i \delta=
\bar{\mathbf{e}}_i'T_i'\Psi_iT_i\bar{\mathbf{e}}_i$.

Using Lemma~\ref{Lem1}, we conclude that
\begin{eqnarray}
  \dot V_i &\le &
-2\alpha_i V_i(e_i) + 2e_i'Y_i^{-1}\dot e_i
  +e_i'(2\alpha_iY_i^{-1}+S_i)e_i \nonumber \\
&&-e^{-2\alpha_i\tau_i}e_i'(t-\tau_i)S_ie_i(t-\tau_i)
+\tau_i ^2\dot
   e_i(t)' R_i\dot e_i(t) \nonumber \\
&& - \bar{\mathbf{e}}_i'\bar \Psi_i\bar{\mathbf{e}}_i;
\label{dotVi.1}
\end{eqnarray}
Then, the statement of the lemma follows from the definition of the matrix
$\tilde\Psi$ and inequality (\ref{dotVi.1}).

\newcommand{\noopsort}[1]{} \newcommand{\printfirst}[2]{#1}
  \newcommand{\singleletter}[1]{#1} \newcommand{\switchargs}[2]{#2#1}

\end{document}